\documentclass[preprint,floats,twoside,tightenlines,eqsecnum,aps,prd, 
showpacs,showkeys,floatfix]{revtex4} 
\usepackage{graphicx}
\usepackage{bm}
 \usepackage{epsfig}

\newcommand{\beqa} {\begin{eqnarray}} 
\newcommand{\eeqa} {\end{eqnarray}} 
\newcommand{\nn} {\nonumber}

\newcommand{\lb}{\label}

\newcommand{\GeV}{\rm GeV} 
\newcommand{\beq}{\begin{equation}} 
\newcommand{\enq}{\end{equation}} 
\newcommand{\beqast}{\begin{eqnarray*}} 
\newcommand{\enqa}{\end{eqnarray}} 
\newcommand{\enqast}{\end{eqnarray*}}

\newcommand{\al}{\alpha} 
 
\newcommand{\ga}{\gamma} 
 
\newcommand{\ep}{\epsilon}

\newcommand{\rh}{\rho} 
\newcommand{\si}{\sigma}

\newcommand{\om}{\omega}

\newcommand{\Ga}{\Gamma}

\begin{document}
\title{  Exclusive electroproduction of $J/\psi$ mesons}
\author{H.G. Dosch }
\affiliation{Institut f\"ur Theoretische Physik, Universit\"at Heidelberg\\
 Philosophenweg 16, D-6900 Heidelberg, Germany }
\author{E. Ferreira}
\affiliation{Instituto de F\'{\i}sica, Universidade Federal do Rio de
Janeiro \\
C.P. 68528, Rio de Janeiro 21945-970, RJ, Brazil   }

\date{\today}

\begin{abstract}
\noindent
A nonperturbative calculation of elastic electroproduction 
of the $J/\psi$ meson is presented and compared to the experimental
data. Our model describes well the observed 
dependences of the cross sections on the photon virtuality $Q^2$ and on 
the energy, and the measured ratio  $R=\sigma_L/\sigma_T$ of 
longitudinal to transverse cross sections.
 \end{abstract}
 \bigskip

 \maketitle

\section{Introduction}
In a previous paper \cite{DF02} we have investigated photoproduction of 
heavy mesons off protons, finding good agreement between experiment and 
our calculations based on a nonperturbative approach to high energy 
scattering \cite{DFK94,DGKP97}, where  no free parameters had to be 
introduced. 
 In this letter we present the results  for electroproduction of 
$J/\psi$ mesons, and  compare our results with the published HERA data. 

In the perturbative approach to electroproduction of heavy mesons the 
coupling of the exchanged gluons to the heavy vector mesons is treated 
perturbatively, while for the coupling of the gluons to the proton an 
external nonperturbative quantity, the gluon density in the proton,
has to be introduced. In our approach  to photo- and electroproduction 
of heavy mesons, the small system, the $\ga ^*-J/\psi$ transition overlap, 
and the large system, the proton, are treated on the same footing and 
the treatment of high energy scattering is based on functional integrals 
\cite{Nac91,DFK94}, which are  evaluated in a nonperturbative approach to 
QCD, the stochastic vacuum model\cite{Dos87,DS88}. 
It has been shown \cite{SSDP02} 
that this nonperturbative approach also implies factorisation in the 
sense that genuine nonperturbative effects in the small system can be 
absorbed into the large system, that is the proton in the present 
case. Although our nonperturbative  approach  is more model dependent 
than the perturbative calculation, it offers the advantage that purely 
hadronic reactions are described with the same set of parameters as photo 
and electroproduction processes. Therefore in our calculations of heavy 
vector meson production no new parameters are introduced and the 
Regge model can be applied directly. Furthermore the  influence 
of confinement effects in the small system can be studied. The WKB 
framework underlying our model also allows to calculate the 
dependence of the production processes on (moderate) momentum 
transfer, which is not directly possible in the perturbative 
approach.

\section{Basic formul\ae~ and general results }
For  convenience we present here  some basic formulae developed in
our previous work on photoproduction \cite{DF02}, where  motivation 
and details can be found. 

The electroproduction amplitude of vector mesons is written 
\beq 
T_{\gamma^* p \to V p,\lambda}(t)= \int d^2 R_1 dz_1  
\psi_{V\lambda}(z_1,R_1)^*\psi_{\ga^* \lambda}(z_1,R_1,Q^2)
J(\vec q,z_1,\vec R_1) ~ , 
\label{int} \enq
with
\beq 
J(\vec q,z_1,\vec R_1)= \int  d^2 R_2 d^2 b \, e^{-i \vec q.\vec b} 
|\psi_p(R_2)|^2  S(b,z_1,\vec R_1,z_2=1/2,\vec R_2) ~ .  
\label{int2} \enq
Here $S(b,z_1,\vec R_1,1/2,\vec R_2)$ is the scattering amplitude of 
two dipoles with separation vectors $\vec R_1,~\vec R_2$, colliding 
with impact parameter vector  $\vec b$; $\vec q$ is the momentum 
transfer of the reaction
\beq
t = -\vec
q\,^2 - m_p^2 (Q^2+M_V^2)/s^2 + O(s^{-3})\approx -\vec q\,^2 ~ .
\enq 

The differential cross section is given by 
\beq
\frac{d \si}{d|t|} = \frac{1}{16 \pi s^2} |T|^2 ~ . 
\enq

The wave functions of the photon and vector meson have been discussed 
extensively in (\cite{DF02}), where it has been shown that the results 
obtained with different forms of the meson wave function are very similar. 
We  use here only the Bauer-Stech-Wirbel (BSW) \cite{BSW87} type 
of wave  function, which is of the general form 
\beq \label{wavegen}
\psi_{V\lambda}(z_1,R_1)=f(z_1)~ \exp [-\om^2 R_1^2/2]\times 
      \mbox{helicity dependent factors} ~ ,
\enq
where $f(z_1)$ contains the normalisation constant and
the $z_1$-dependence of the meson wave function, with the form  
\beq
f(z) = \frac{N}{\sqrt{4 \pi~}} ~\sqrt{z(1-z)~}~ 
\exp\Big[-\frac{M_V^2}{2 \om^2}(z-\frac{1}{2})^2\Big] ~ .
\enq

 The size of the vector meson is determined by $\om$, 
which is fixed by the leptonic decay width of the meson. 

 After summation over helicity indices, the overlaps of the
photon and vector meson wave functions are given by 
 \beqa 
 \lefteqn{\rho_{\gamma^*, V,\pm 1}(z_1,R_1,Q^2)=
\psi_{V, \,\pm 1}(z_1,R_1)^*\psi_{\ga^*, \,\pm 1}(z_1,R_1,Q^2)  = }\nn\\
&&\hat e_f\frac{\sqrt{6\alpha}}{2\pi}  
   ~~f(z_1) \exp[-\om^2 R_1^2/2]\nn \\ 
&&\times \Big( \ep ~\om^2 R_1 \big[z_1^2+(1-z_1)^2\big] K_1(\ep~R_1) 
   + m_f^2 K_0(\ep ~R_1)\Big) \nn \\
\lefteqn{
\rho_{\gamma^*,V,0}(z_1,R_1,Q^2)=\psi_{V,\,0}(z_1,R_1)^*\psi_{\ga^*,\,0}
(z_1,R_1,Q^2)=}\\
&&- 16 \, \hat
e_f\frac{\sqrt{3\alpha}}{2\pi}   ~\om~f(z_1) \exp[-\om^2 R_1^2/2] 
z_1^2(1-z_1)^2 Q K_0(\ep ~R_1) ~  . \nn \enqa
Here $$\ep=\sqrt{z_1(1-z_1) Q^2+m_f^2} ~ ,$$
$K_0$, $K_1$ are the modified Bessel functions, $\lambda=\pm 1$ 
and $0$  denote transverse and longitudinal polarisations 
of the vector meson and the photon, $m_f$ is the quark mass and  
$\hat e_f$ is the quark charge in units of the elementary charge 
for each flavour $f$. All  parameters are fixed from other processes 
(see \cite{DF02}) and   all observables can be calculated from 
(\ref{int}).

  The energy dependence in our model is motivated by  
the two-pomeron model of Donnachie and Landshoff \cite{DL98}.   For  
$R_1 \leq r_c \approx 0.22$ fm the coupling through  the hard 
pomeron induces the energy dependence  $(s/s_0)^{0.42}$,  while the 
coupling of large dipoles follows the soft pomeron energy dependence 
 $(s/s_0)^{0.0808} $. The reference energy is $s_0=(20~  \GeV)^2 $.
We therefore split the integration over $R_1$  appearing  in (\ref{int}) 
into two parts, as fully described in our study of $J/\psi$ 
photoproduction \cite{DF02}.

 Before proceeding to the comparison with experimental data we wish
to present some general approximate features of our model which
describe the overall situation quite well and provide a background 
against which finer details can be studied. 

If $R_1$ is small compared to the extension of the proton and 
$|t|=\vec q\,^2 <1 $ GeV$^2$,   we obtain after integration over 
the azimuthal angle of $\vec R_1$, the simple expression for the 
amplitudes
\beq \label{simple}
T^\lambda _{\gamma^* p \to V p}(t )\approx (- 2 i s) 2 \pi  \frac{C
e^{-a|t|}}{1+b|t|} \int d R_1 dz_1 \, R_1^3 e^{-\eta R_1^2 |t|}
\rh_{\gamma^*,V,\lambda}(z_1,R_1,Q^2) ~ .   \enq
For the case of $J/\psi$ production the numerical values for the 
constants in (\ref{simple}) are 
\beq
\label{parameters}
C=2.28,~~~a=0.847 \mbox{ GeV}^{-2},~~~b=3.85 \mbox{ GeV}^{-2} ,
~~~\eta=0.059 ~ .
\enq

This result  means that the integrated expression in (\ref{int2})
is  proportional to $R_1^2 \exp[-\eta R_1^2 |t|]$ and that the 
main t dependence in the amplitude is factorized out,  
being independent  of $Q^2$ and of $W$. 

Since in the overlap functions with heavy vector mesons 
$\eta R_1^2 $ is very small
, the exponential factor $e^{-\eta R_1^2 |t|}$ can be extracted from 
the integral in (\ref{simple}), and represented by an external factor  
$e^{-\eta R_{m}^2 |t|}$, where  $R_{ m}$ is an  appropriate mean  value 
of $R_1$.  The numerical calculation shows that a very good 
representation for  the $Q^2$ dependence of this external factor 
is given by $R_m ^2\approx 12/({Q^2 +M_V^2})$ . 

 We can then  write for the scattering amplitude 
\beqa \label{simple_s}  T^\lambda _{\gamma^* p \to V p}(s,
t;Q^2 )&\approx&(- 2 i s)  \frac{C e^{-a|t|}}{1+b|t|} 
 \exp \bigg[-12 \eta |t|/(Q^2 + M_V^2)\bigg]    \nn \\ 
&\times& \left(
T_h^\lambda(Q^2)\left(\frac{s}{s_0}
\right)^{\ep_h}  +T_s^\lambda(Q^2)\left( \frac{s
}{s_0}\right)^{\ep_s}\right) \enqa
with 
\beqa \lb{reduced}
T_h^\lambda(Q^2)&=&2 \pi \int_0^{r_c} d R_1 dz_1 \, 
\left(\frac{R^2}{r_c^2}\right)^{\ep_h}
R_1^3 ~ \rh_{\gamma,V,\lambda}(Q^2,z_1,R_1) ~ ,  \nn \\ 
T_s^\lambda(Q^2)&=&2 \pi \int_{r_c}^\infty dR_1 dz_1 \, 
R_1^3 ~ \rh_{\gamma,V,\lambda}(Q^2, z_1,R_1)   ~ .
\enqa
The parameters for the energy dependence are  
$s_0=(20\mbox{ GeV})^{2},~~\ep_h=0.42,~~\ep_s=0.0808 $~ . 

\begin{figure}[ht]
\vskip 2mm
\includegraphics[height=10cm,width=10cm]{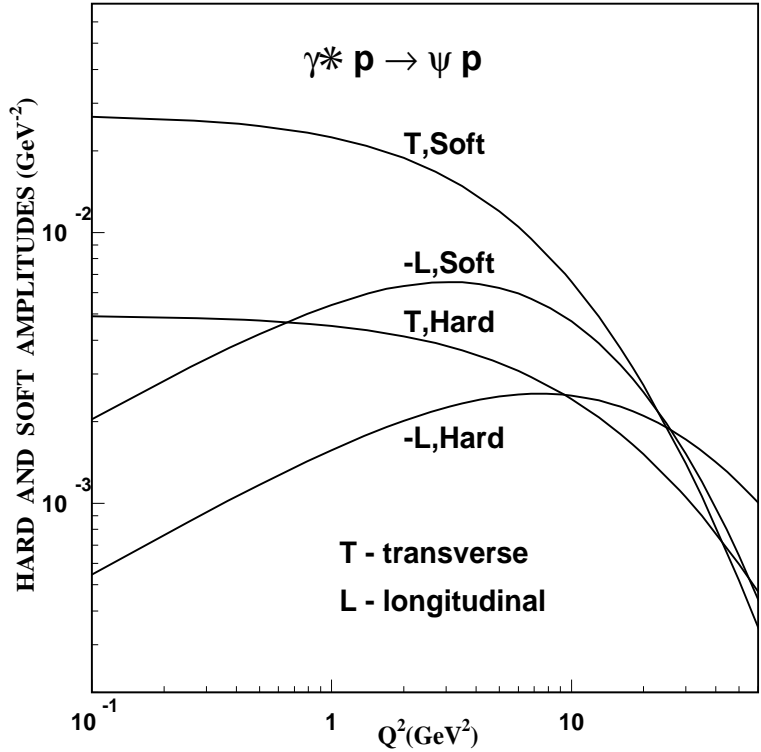}
\caption{\label{parts} Hard and soft parts of the reduced amplitudes
for transverse and longitudinal polarisations, according to 
Eqs. (\ref{reduced}). These functions can be  represented  by the 
parametrisations (\ref{para1}),(\ref{para2}).}
\end{figure}

Eq. (\ref{simple_s}) allows us to express all observables through the
functions $T_h^\lambda(Q^2)$ and $T_s^\lambda(Q^2)$ (\ref{reduced}).  
We display these amplitudes in Fig. \ref{parts}. 
They can be parametrized for the range $0 \leq Q^2 \leq 100$ GeV$^2$ 
in the forms   
\beqa\label{para1}
T_s^{\pm 1}(Q^2)&=& 0.0267/(1+\frac{Q^2}{15.03})^{2.71} ~ , \nn \\
T_h^{\pm 1}(Q^2) &=& 0.00488/(1+\frac{Q^2}{20.46})^{1.72} ~ ,
\enqa
\beqa\label{para2}
T_s^{0}(Q^2)&=& -0.00635 Q /(1+\frac{Q^2}{18.00})^{3.23} ~ , \nn \\ 
T_h^{0}(Q^2) &=& -0.00170 Q /(1+\frac{Q^2}{20.61})^{1.9} ~ ,
\enqa
where $Q^2$ is measured in GeV$^2$. 

The overall  $|t|$ dependence determined by the form factor
$$ F(|t|)= \frac{ e^{-a|t|}}{1+b|t|} \exp[-12 \eta
|t|/(Q^2 + M_V^2)] $$
yields  a characteristic curvature in the log plot. 
 We can introduce an effective slope  defined as  
\beq
B=\bigg[\frac{d \sigma}{d|t|}\bigg]_0 ~ / \int d|t| \frac{d \sigma}{d|t|}
\enq
which is the inverse of the integral  $\int F(|t|)^2  d|t|$
and  varies monotonously from $B=7.038 \mbox{ GeV}^{-2}$ to 
$ B= 6.839 \mbox{ GeV}^{-2} $ in the interval $0 \leq Q^2 < \infty$. 

In any approach where the wave function of the heavy meson is
taken into account \cite{Predazzi97,Hufner00}
nonperturbative quantities like the  ``meson radius'' $1/\om$ 
enter even at the ``small'' side of the 
 interaction which determines the ``hard'' scale. It is therefore
 interesting to study  the strictly non-relativistic limit of the 
meson  wave function, where the ratio $\om/M_V$ goes to zero and 
the meson  radius drops out in the final expressions for the amplitudes. 
In this  limit, which also
implies  $m_f \to  M_V/2$,  all integrals can be performed 
analytically if we neglect the energy dependence by putting 
$\ep_s=\ep_h=0$.
We  obtain the asymptotic results
\beqa \label{nonrel} T^{\pm1} _{\gamma^* p \to V p}(t,Q^2)/(-2 i s) 
   &\approx& \frac{C e^{-a|t|}}{1+b|t|}
 \frac{\sqrt{3}}{\sqrt{\al}} \frac{16}{(Q^2 + M_V^2)^2}
M_V^{3/2} \Ga_{e^+e^-}^{1/2} ~ ,    \nn \\ 
T^0_{\gamma^* p \to V p}(t,Q^2) &\approx& \frac{Q}{M_V}T^{\pm1
}_{\gamma^* p \to V p}(t,Q^2) ~ .  \enqa

\begin{figure}[ht]
\vskip 2mm
\includegraphics[height=10cm,width=10cm]{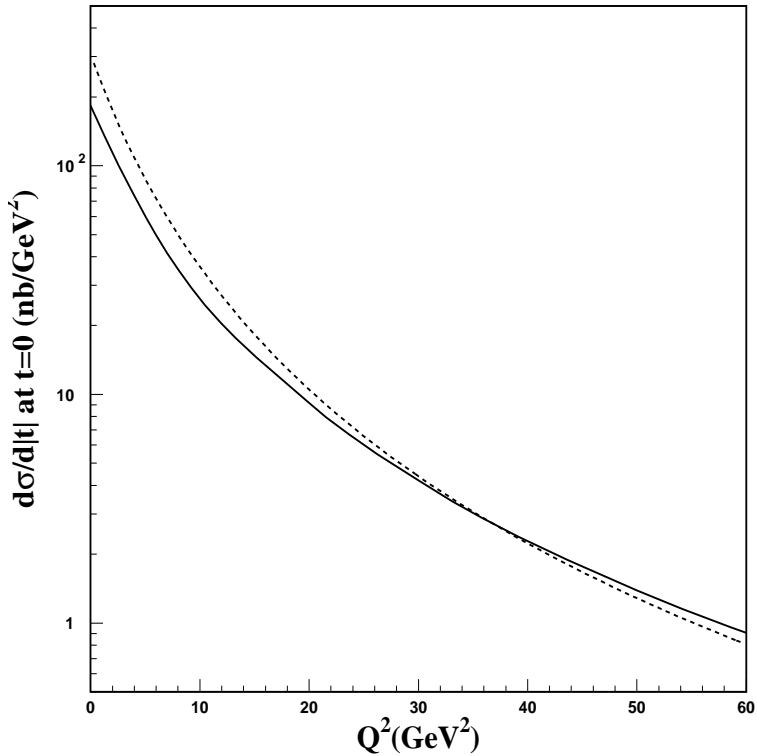}
\caption{\label{forward} Influence of the nonperturbative 
parameters of the $J/\psi$ meson on the  $Q^2$ dependence
of the forward differential cross section. In solid line 
our calculations without energy dependence  ($\ep_h=\ep_s=0$)
and in dashed line the asymptotic form of Eq. (\ref{nonrel}).  } 
\end{figure}

In Fig. \ref{forward} we show in solid line the full result 
(with $\ep_h=\ep_s=0$) and in dashed line the asymptotic form of 
(\ref{nonrel}) for forward  differential 
cross section  $\bigg[d\si/d|t|\bigg](t=0)$.

\section{Numerical results and comparison with experiments}
   
\begin{figure}[ht]
\vskip 2mm
\includegraphics[height=10cm,width=10cm]{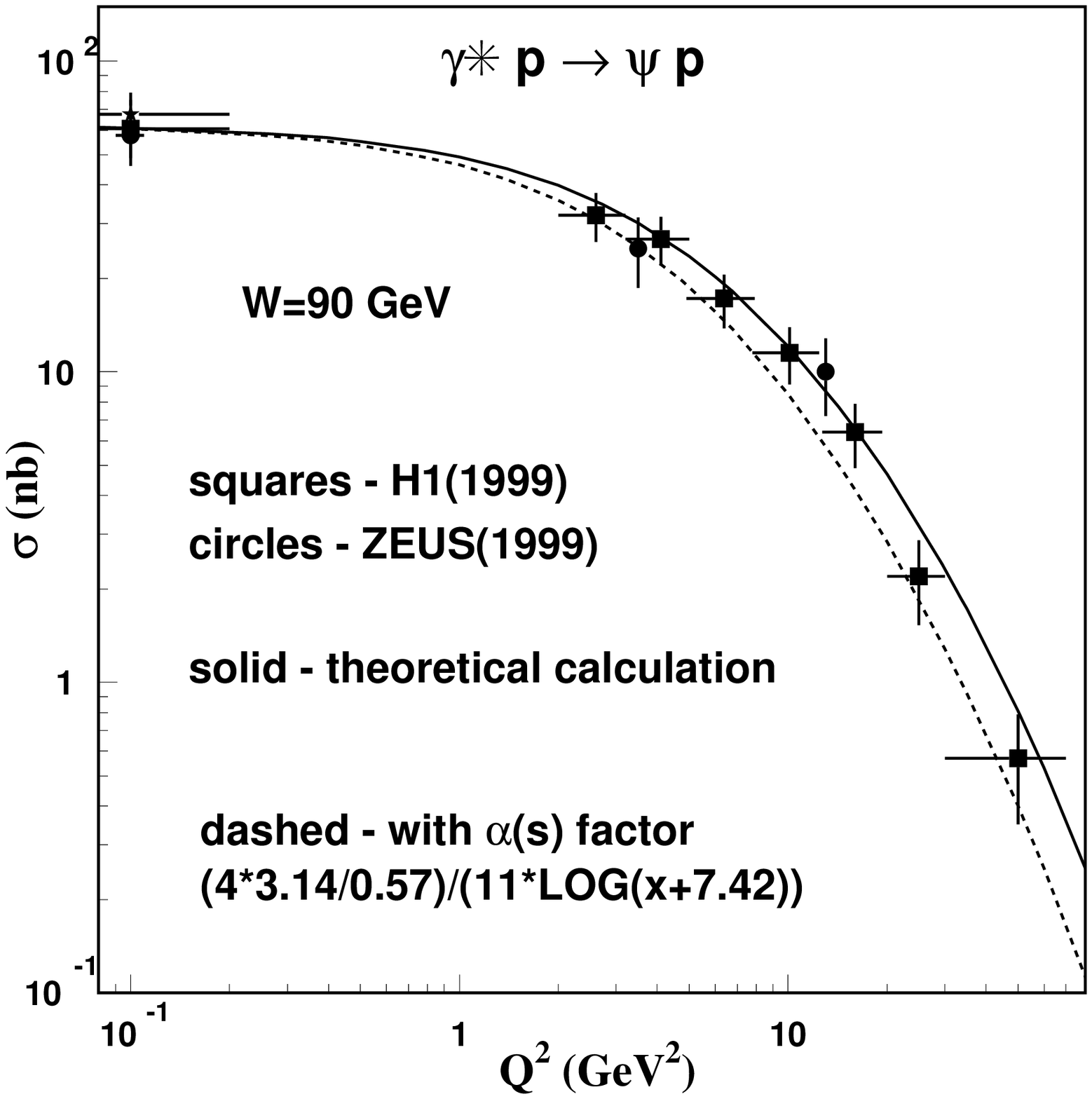}
\caption{\label{pure} $Q^2$  dependence of the integrated elastic 
cross section  $\sigma=\sigma_T+\sigma_L$ at the energy $ W=90~  \GeV $, 
compared to data from Zeus \cite{Zeus99} (circles) and H1 \cite{H199} 
(squares). The full line represents our results as explained in the 
text. In the dashed line an extra factor is introduced to account for 
the possible effect of a $Q^2$ dependence of the running strong 
coupling constant \cite{DD02}. } 
\end{figure}

Our numerical calculations presented in the figures below  with 
experimental data are  
exact evaluations of the model, based on Eqs.(\ref{int},\ref{int2}). 
However, we stress that  
 Eqs.(\ref{simple_s},\ref{reduced},\ref{para1},\ref{para2})
represent very accurately   these results.

In Fig. \ref{pure} we show the integrated  $J/\psi$ production 
cross section $\sigma=\sigma_T+\sigma_L$ at $W=90~ \GeV$ as a function 
of $Q^2$, together with the HERA data 
from the H1 and Zeus collaborations \cite{Zeus99,H199}. The solid 
line corresponds to the exact equations (\ref{int},\ref{int2}), 
while  for the  dashed line there is  a multiplying factor  
$$\frac{4\pi/0.57}{11 \log (Q^2+7.42)}$$  introduced \cite{DD02} 
to take into account the running of the strong coupling in the model 
which otherwise is purely nonperturbative. 

\begin{figure}[ht]
\vskip 2mm
\includegraphics[height=10cm,width=10cm]{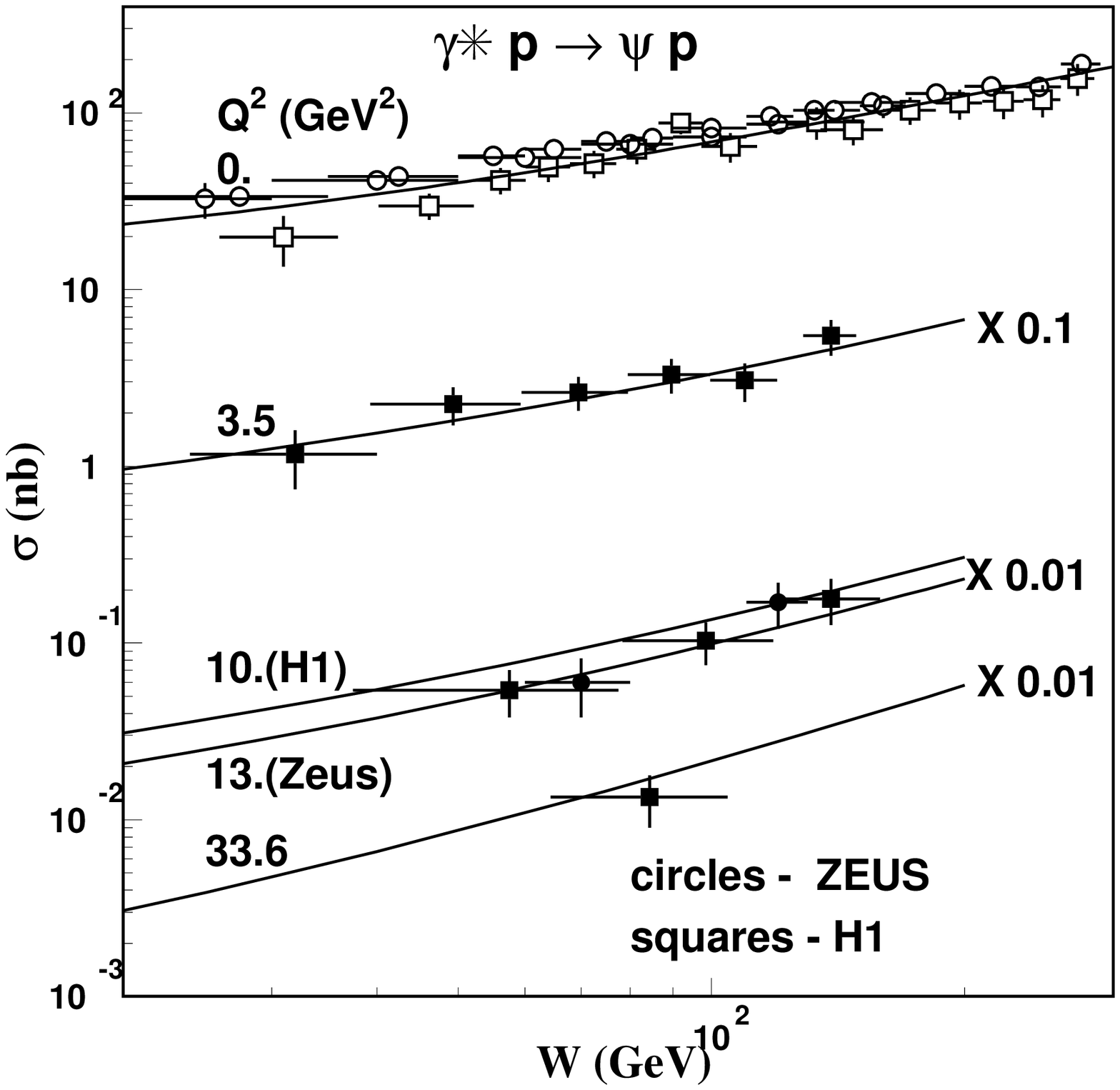}
\caption{\label{energy} Energy dependence of the integrated cross 
section for several values of $Q^2$ compared to data from Zeus in 
full \cite{Zeus99} and empty \cite{Zeus02} circles and from H1 in 
full \cite{H199} and empty \cite{H100} squares. } 
\end{figure}

The energy dependence adopted in our model is fully compatible with 
the existing published data \cite{Zeus99, H199,H100,Zeus02} as can 
be seen in Fig.\ref{energy}. We repeat in this figure our results for 
photoproduction (empty circles and squares) which have been shown 
before \cite{DF02} to agree well with the data. 

The energy dependences of the cross section  have  general forms 
$$ \sigma_{T,L}(Q^2)=(A^{T,L}_{\rm hard}(Q^2)\,W^{2 \epsilon_h} 
+A^{T,L}_{\rm soft}(Q^2) \,W^{2 \epsilon_s})^2 ~ , $$ 
where $A^{T,L}_{\rm hard}(Q^2)$ and $A^{T,L}_{\rm soft}(Q^2)$ can 
be easily 
calculated from (\ref{simple_s}), (\ref{para1}) and (\ref{para2}). 

Often experimental cross sections are fitted through a single power
$W^{\delta(Q^2)}$, with $W=\sqrt{s}$, which must be used 
in a  limited energy range.  For small $Q^2$ the soft pomeron is 
important and $\delta $ is small, but for larger $Q^2$ the hard 
pomeron becomes dominant and the  exponent
$\delta \equiv 4 \ep_{\rm eff} \to 4 \epsilon_{\rm h}= 0.168$ as  
$Q^2 \to \infty$.  A convenient interpolation formula in the range 
$20 \leq W \leq 200$ GeV is 
\beq \label{delta}
\ep_{\rm eff}(Q^2)=\ep_{\rm h}-\frac{1}{(3.28+Q^2/M_V^2)^{1.22}} ~ .
\enq  

\begin{figure}[ht]
\vskip 2mm
\includegraphics[height=10cm,width=10cm]{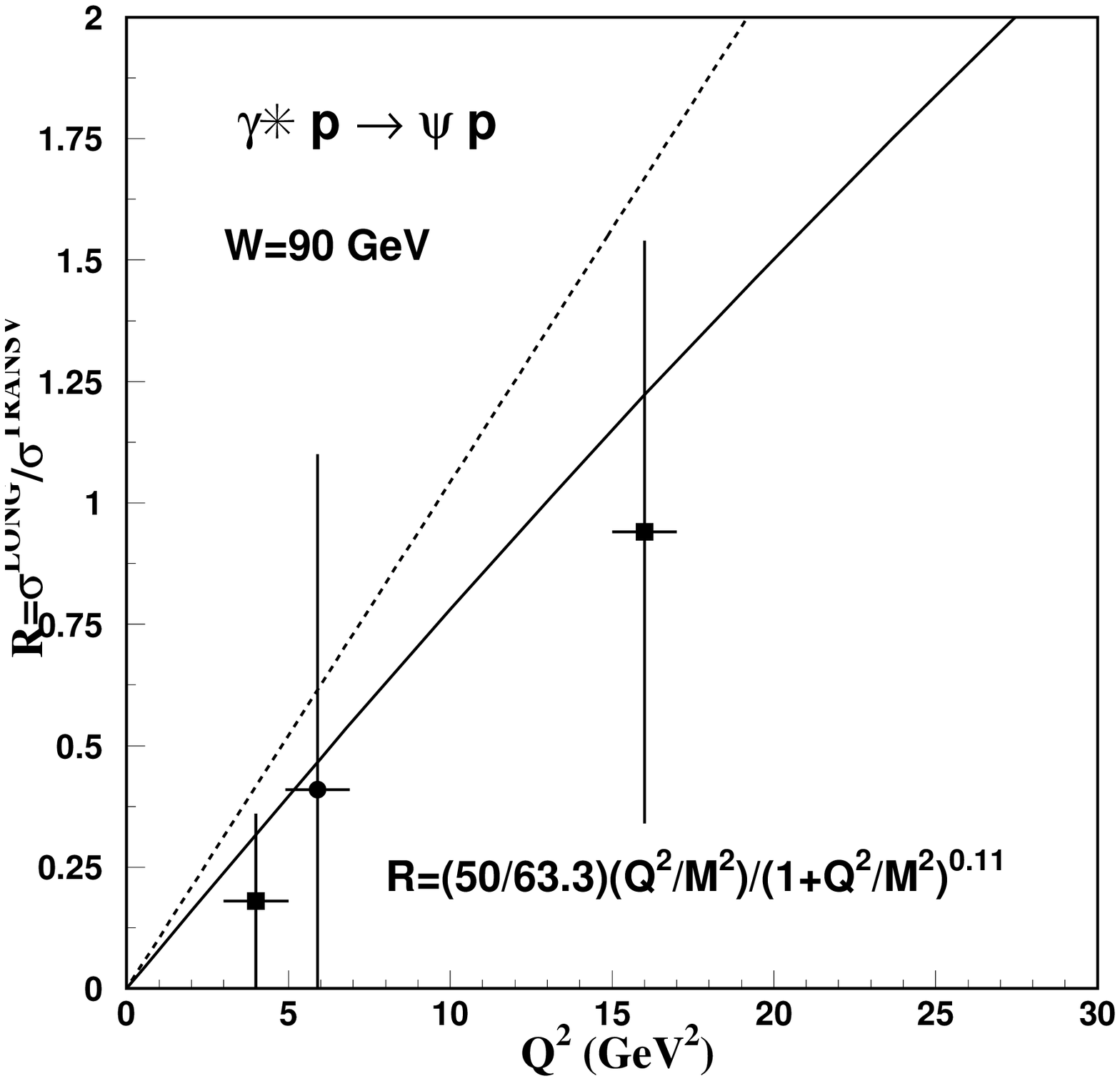}
\caption{\label{ratio2} Ratio between longitudinal and transverse
cross sections at $W=90~ \GeV $ as a function of $ Q^2 $ compared 
 to the data from Zeus \cite{Zeus99} (circle) and H1 \cite{H199}
(squares) collaborations. The solid line is our theoretical result, 
for which we give a parametrization. The dashed line represents the 
asymptotic calculation (size zero for the vector meson). }
\end{figure}

Our results for the transverse and longitudinal cross sections at 
$W=90~  \GeV $ can be parametrised in the forms
\beq
\si_T(Q^2) = \frac{63 }{(1 + Q^2/M_{J/\psi}^2)^{3.17}}\mbox{ nb}
  ~~~ {\rm and} ~~~
\si_L(Q^2) = \frac{50 ~  Q^2/M_{J/\psi}^2 }
            {(1 + Q^2/M_{J/\psi}^2)^{3.28}} \mbox{ nb} ~ .
\enq
The ratio $R=\sigma_L/\sigma_T$ of longitudinal to transverse 
cross section is displayed in Fig. \ref{ratio2}. The solid line 
shows our result for $W=90$ GeV, which agrees  well with the data 
\cite{Zeus99,H199} at the same energy. The dashed line  is the 
asymptotic result $R_{\rm asymp} = Q^2/M_{J/\psi}^2$, from 
(\ref{nonrel}).

The differential cross section $d\si/d|t|$ and its energy dependence
obtained from our model have been shown \cite{DF02}  to describe well 
the data for $J/\psi$ photoproduction in a wide energy range. There 
are no published data of $d\si/d|t|$  for electroproduction  to be 
compared to our calculations, which predict that the shape of the 
angular distribution is almost independent of  $Q^2$.

\section{Conclusions}
Our nonperturbative model describes the data well. We found that 
the main features are expressed in the simplified equations
 (\ref{simple_s},\ref{reduced},\ref{para1},\ref{para2}) which 
represent very well the full results of our model. This is due to 
the small range  of  the overlap functions  compared to the proton 
size and to the range of  the  correlation functions of the QCD 
vacuum , which  determine  the interactions in our model.
 As a consequence of these properties of the amplitudes, in our 
 nonperturbative treatment of the electroproduction of heavy  
 vector mesons  (here with particular application to $J/\psi$ 
production) some quite general features emerge, which are mainly a
consequence  of the general approach to high energy scattering 
~ \cite{Nac91,DFK94,DGKP97}  and  not of the specific stochastic
vacuum model \cite{Dos87,DS88}, which  yields the   numerical
 values of the parameters in (\ref{parameters}).

   The main  shape of the $Q^2$ dependence is determined by the 
overall overlap strength (\ref{reduced}). Our 
calculations, which reproduce well the data for $J/\psi$ production 
 for $Q^2 \leq 60~ \GeV^2$  deviate  considerably from the asymptotic 
result (\ref{nonrel}) and shows that genuine nonperturbative 
parameters, like the size of the vector meson ($1/\om$), have a 
considerable influence even on the ``hard part'' of the interaction.  

Fig. \ref{energy} shows that the $W$-dependence based on the 
two-pomeron 
model \cite{DL98} is appropriate to describe the existing data. 

In the ratio $ R=\sigma_L/\sigma_T $  the characteristic numbers 
coming from the stochastic vacuum model nearly cancel  out, 
as can be seen in (\ref{simple}), and the ratio  depends 
solely on the overlap strengths defined in (\ref{reduced}). 
Due to the energy dependence induced by the overlap function, see
(\ref{simple_s},\ref{reduced}), the ratio will also exhibit a 
characteristic  energy dependence.   At $W=90 ~ \GeV $ 
the  ratio agrees reasonably well with the data shown in 
Fig. \ref{ratio2}. It is important to  check this result against 
future data at other energies and for a wider range of $Q^2$. 

The predicted  $t$ dependence is nearly
universal, namely  independent of $Q^2$ and $W$, with  only small
corrections due to the term $e^{-\eta R_1^2|t|}$ in (\ref{simple}) 
and  slight  dependence on $W$ due to the different energy factors 
controlling the couplings of small and large dipoles.

\begin{acknowledgments} 
The authors wish to thank DAAD(Germany), 
CNPq(Brazil), CAPES(Brazil) and  FAPERJ(Brazil) for support of the 
scientific collaboration program between Heidelberg, Frankfurt  and 
Rio de Janeiro groups working on hadronic physics. 
\end{acknowledgments}


\begin{thebibliography}{99}
\bibitem{DF02}
H.G.Dosch and E.Ferreira,
Eur.\ Phys.\ J.\ C {\bf 29} (2003) 45
\bibitem{DFK94}
H.G. Dosch, E. Ferreira and A. Kramer,
Phys.\ Rev.\ D {\bf 50} (1994) 1992
\bibitem{DGKP97} 
H.~G.~Dosch, T.~Gousset, G.~Kulzinger and H.~J.~Pirner, 
Phys.\ Rev.\ D {\bf 55} (1997) 2602 

\bibitem{Nac91}
O.~Nachtmann,
Annals Phys.\  {\bf 209} (1991) 436.

\bibitem{Dos87}
H.~G.~Dosch,
Phys.\ Lett.\ B {\bf 190} (1987) 177.

\bibitem{DS88}
H.~G.~Dosch and Y.~A.~Simonov,
Phys.\ Lett.\ B {\bf 205} (1988) 339.


\bibitem{SSDP02}
A.~I.~Shoshi, F.~D.~Steffen, H.~G.~Dosch and H.~J.~Pirner,
Phys.\ Rev.\ {\bf D66}: 094019 (2002)

\bibitem{BSW87}
M.~Bauer, B.~Stech and M.~Wirbel,
Z.\ Phys.\ C {\bf 34} (1987) 103.

\bibitem{DL98}
A.~Donnachie and P.~V.~Landshoff,
Phys.\ Lett.\ B {\bf 437} (1998) 408

\bibitem{Predazzi97}
J. Nemchik, N.N. Nikolaev, E. Predazzi and B.G. Zakharov, 
Z.\ Phys.\ C {\bf 75}, 71  (1997) 

\bibitem{Hufner00}
J. H\"ufner, Yu.P. Ivanov, B.Z. Kopeliovich and A.V. Tarasov,
Phys.\ Rev.\ D {\bf 62}, 094022 (2000)

\bibitem{Zeus99}
J.Breitweg et al., Zeus Coll., 
Eur.\ Phys.\ J.\ C {\bf 6} (1999) 603  

\bibitem{H199}
C. Adloff et al., H1 Coll.
Eur.\ Phys.\ J. C {\bf 10} (1999) 373 

\bibitem{DD02}
A.~Donnachie and H.~G.~Dosch,
Phys.\ Rev.\ D {\bf 65} (2002) 014019

\bibitem{H100}
C. Adloff et al., H1 Coll.
Phys.\ Lett.\ B {\bf 483} (2000) 23

\bibitem{Zeus02}
S. Chekanov et al., Zeus Coll.,
Eur.\ Phys.\ J.\ C {\bf 24} (2002) 345

\end{thebibliography}
\end{document}